# A Simple, Fast and Fully Automated Approach for Midline Shift Measurement on Brain Computed Tomography

Huan-Chih Wang, Shih-Hao Ho, Furen Xiao, Jen-Hai Chou

2 Mar 2017

***Abstract:*** Brain CT has become a standard imaging tool for emergent evaluation of brain condition, and measurement of midline shift (MLS) is one of the most important features to address for brain CT assessment. We present a simple method to estimate MLS and propose a new alternative parameter to MLS: the ratio of MLS over the maximal width of intracranial region (MLS/ICW$_{MAX}$). Three neurosurgeons and our automated system were asked to measure MLS and MLS/ICW$_{MAX}$ in the same sets of axial CT images obtained from 41 patients admitted to ICU under neurosurgical service. A weighted midline (WML) was plotted based on individual pixel intensities, with higher weighted given to the darker portions. The MLS could then be measured as the distance between the WML and ideal midline (IML) near the foramen of Monro. The average processing time to output an automatic MLS measurement was around 10 seconds. Our automated system achieved an overall accuracy of 90.24% when the CT images were calibrated automatically, and performed better when the calibrations of head rotation were done manually (accuracy: 92.68%). MLS/ICW$_{MAX}$ and MLS both gave results in same confusion matrices and produced similar ROC curve results. We demonstrated a simple, fast and accurate automated system of MLS measurement and introduced a new parameter (MLS/ICW$_{MAX}$) as a good alternative to MLS in terms of estimating the degree of brain deformation, especially when non-DICOM images (e.g. JPEG) are more easily accessed.

## INTRODUCTION

Brain computed tomography (CT) is indispensable for evaluation of brain abnormality, especially in acute settings such as traumatic brain injury (TBI) and spontaneous intracerebral hemorrhage (ICH). It is crucial to provide appropriate care in a timely fashion, a fast and accurate interpretation of the brain CT images is thus the first key step of achieving a better neurological prognosis. Through the Monro-Kellie hypothesis, space-occupying lesions (e.g. hematoma, tumor, abscess, etc.) can cause local or global brain shifts, followed by herniation, changes in intracranial pressure (ICP), brainstem compression and death. Since the human head is a roughly symmetric structure, observation on the distortion of midline anatomy, or better known as "midline shift (MLS)", serves as a convenient feature for prediction of increased ICP [1, 2]. Although the visual inspection and manual measurement of MLS has already become a standard and routine task for experienced physicians (especially specialists frequently encountered with interpretation of brain CT images, including neuroradiologists, neurosurgeons and neurologists), accurate brain CT interpretation might still be challenging for other health care providers. Emergency physicians are usually the first ones to be involved in the management of patients with TBI and spontaneous ICH, but the reliability of their brain CT reading is often questioned [3, 4]. Through the advancement of computational technologies, we believe that computer-aided diagnosis may have the potentials to improve the accuracy and processing speed of both non-experts and experts in terms of brain CT interpretation.

It is traditionally suggested that the degree of MLS should be measured at the level of foramen of Monro where the septum pellucidum serves as the true midline [2]. It is also often measured as the distance of the septum pellucidum to the ideal midline (Figure 1) in daily practices. The latter measuring method is adopted in several automated or semi-automated methods of measuring the MLS, which are discussed in previous

H. C. Wang (b88401030@ntu.edu.tw) is with the Division of Neurosurgery, Department of Surgery at National Taiwan University Hospital Hsin-Chu Branch, Hsin-Chu, Taiwan.
F. Xiao (xfuren@gmail.com) is with the Division of Neurosurgery, Department of Surgery at National Taiwan University Hospital, Taipei, Taiwan
S. H. Ho (shho@deep01.com) and J. H. Chou (david@deep01.com) are with Deep01, Ltd.
Please address correspondence to J. H. Chou.

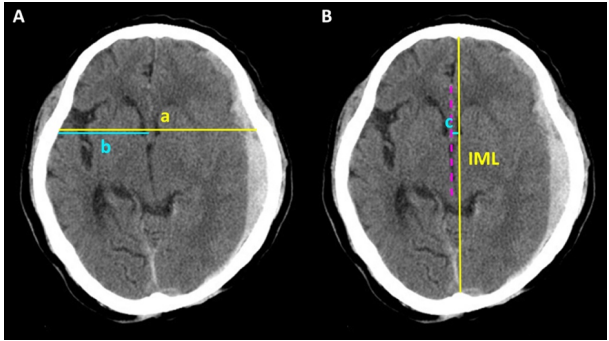

**Figure 1. Midline shift measurement**
(A) At the level of foramen of Monro, the intracranial length ("a") and the distance from the skull to the septum pellucidum ("b") are measured. The midline shift can then be calculated as: (a/2) – b.
(B) Another commonly adopted approach for midline shift measurement. The ideal midline (IML) is determined as the connection line between the anterior-most point of anterior falx and the posterior-most point of posterior falx. Another parallel line passing through the septum pellucidum is plotted (dotted line), the midline shift can then be measured as the distance between these two lines ("c").

literatures [5-8]. A measured MLS more than 0.5 cm on the initial brain CT images was considered to be significant and predicts poor neurological outcomes (with a positive predictive value of 78%) [9, 10]. Nonetheless, given the variable sizes of human heads among different age groups and different ethnicities, it is reasonable to question that whether an arbitrary absolute value should be applied to all patients. We therefore introduced an alternative approach by using the ratio of the measured MLS divided by intracranial width (ICW) and compared its diagnostic power to that of direct MLS measurement.

**Table 1. List of Admission Diagnosis**

| Trauma (n = 24) | EDH | 1 |
|---|---|---|
| | SDH | 12 |
| | Contusion/ICH | 5 |
| | SAH | 5 |
| | Depressed skull fracture | 1 |
| Non-trauma (n = 19) | Spontaneous ICH | 11 |
| | Spontaneous ICH + IVH | 6 |
| | Brain tumor | 1 |
| | Epidural abscess | 1 |

The list of all 43 patients with CT scan available. The CT images of the brain tumor case and the epidural abscess case were excluded because of the difficult measurements even for human experts.
EDH: epidural hemorrhage; SDH: subdural hemorrhage; ICH: intracerebral hemorrhage; SAH: subarachnoid hemorrhage; IVH: intraventricular hemorrhage.

## MATERIALS AND METHODS

From January 2015 to March 2015, 50 consecutive patients were admitted to the intensive care unit (ICU) of National Taiwan University Hospital Hsin-Chu Branch under neurosurgical service. Seven patients were excluded from this study due to no CT images could be retrieved from the PACS archive. There were 29 males and 14 females amongst the remaining 43 patients, with their ages ranging from 13 to 93 (60.8 ± 22.0) years. Head injury was the cause of ICU admission in 24 patients, while the other 19 patients admitted due to non-traumatic diagnoses. Three patients did not present with intracranial hemorrhage, one of them had a depressed skull fracture, one had a brain tumor and the other had an epidural abscess. The main diagnoses of the 43 patients were listed in Table 1. The brain tumor and the epidural abscess were later excluded due to marked difficulty in measuring the MLS or intracranial width for human experts.

We collected the first non-enhanced brain CT study of each patient (either performed at ER or after admission) and extracted the series containing 5-mm axial images only. A total of 11 sets from these CT images were performed at 7 outside hospitals. The original DICOM images were set to brain window (Hounsfield center 40, width 80) and downloaded to a personal computer as JPEG (8-bit grayscale) format files, with each image containing 512 x 512 pixels. The field of view (FOV) ranged from 20.8 to 28.7 (mean: 24.6 ± 1.5) cm, resulted in a variable resolution of 0.41~0.56 (mean: 0.48 ± 0.03) mm per pixel (equal to the "pixel spacing" parameter in DICOM 3.0).

*Manual measurements:*

Manual measurements were conducted by 3 board-certified neurosurgeons independently. For each set of CT images, the doctors were asked to answer whether there was significant midline shift or not ("first glance evaluation") prior to actual measurement. After this gross evaluation, two parameters were asked to be recorded: the MLS and the maximal intracranial width ($ICW_{MAX}$, the widest intracranial width in the same slice of MLS measurement) in the designated slices where foramen of Monro were clearly visible (Figure 2).

*Computer measurements:*

As previously stated, MLS should be measured at the level of foramen of Monro and thereby leaving only a small subset of axial brain CT slices usable. For a fully automated system, the calibration of head rotation is also an important part of the system. Both appropriate CT slice-selection and the calibration algorithms are not subjects to discuss in this context because it is difficult to quantify the accuracy and efficiency of these two processes. Furthermore, the slice-selection algorithm is not needed in this study as we pre-select the CT slices for both human and computer to measure. Methods to find ideal midline (IML) and MLS estimation are described in the followings.

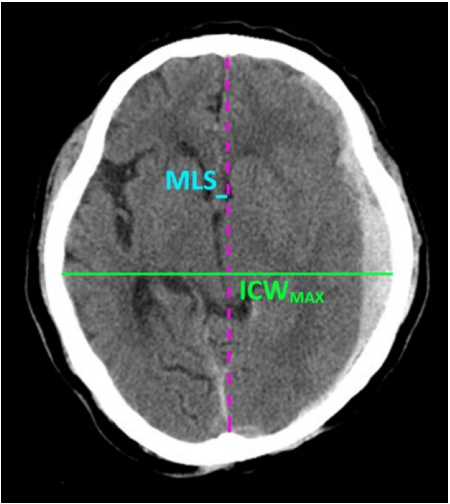

**Figure 2. Brain CT parameters to be measured**
Two parameters were asked to be measured by neurosurgeons and by computer independently: the lengths of MLS and ICW$_{MAX}$. MLS was plotted through the method described in Figure 1B, and ICW$_{MAX}$ is the horizontal line perpendicular to IML (dotted line) which marked the widest diameter in the intracranial region in the same CT image slice.
MLS: midline shift; ICW$_{MAX}$: maximal intracranial width; IML: ideal midline.

Algorithms for automated ideal midline detection were extensively studied previously and various definitions were used to define IML [6, 11-16]. We altered the IML definition to the line passing through the anterior bone protrusion and the center of mass of the intracranial region. This modified definition was chosen for two reasons: simplicity and robustness. Finding the mass center involved naive matrix operations only, which made our algorithm simpler and more efficient. As for robustness, we noticed that there was a falx cerebri detection process for algorithms of intact midline recognition which was sometimes difficult due to physiological and pathological variations. This made detecting the falx much more difficult and sometimes unreliable. Finding the center of mass, on the other hand, caused only very little computing burden and its geometric meaning made our IML passing through the center of the intracranial region.

Our method for MLS estimation was based on a simple observation: the reference points for estimating MLS were relatively dark (mainly the frontal horns of lateral ventricles and the 3rd ventricle) compared with other brain tissues. A new approach was adapted to depict the deformation of brain tissues by calculating a "weighted midline (WML)". This weighted midline was simply a curve connecting a set of weighted averages in each row of the image matrix. We defined the weight factor $w$ for pixel location $x_{ij}$ ($i$ for row and $j$ for column) given its grayscale value $g_{ij}$ (between 0 and 255):

$$w(x_{ij}) = e^{-\alpha g_{ij}} \quad (1)$$

where $w(x_{ij})$ was the (unnormalized) weight of pixel at location $x_{ij}$ and $\alpha$ was a constant; the weighted averages $m$ along rows were then calculated. These weighted averages were essentially a one-dimensional center of mass, (inversely) weighted by the intensity value of each pixel:

$$m(x_i) = \frac{\sum_{j=l_i}^{r_i} x_{ij} w(x_{ij})}{\sum_{j=l_i}^{r_i} w(x_{ij})} \quad (2)$$

where $l_i$ and $r_i$ were the left and right boundaries of row $i$. In our weighting scheme, we gave higher weights for pixels with low intensities (black) and quickly reduce the weights exponentially for pixels with high intensities (white). After the weighted averages were calculated for each row in the intracranial region, smoothing of the results by averaging over its neighbors (convolution) was further carried out. In order to measure MLS, a segment of WML near the foramen of Monro was isolated. MLS was therefore estimated by taking the maximum distance from this segment of WML to the IML (examples were shown in Figure 3).

By definition, ICW$_{MAX}$ was easily generated by finding the maximal horizontal intracranial width (perpendicular to the IML). The ratio of MLS/ICW$_{MAX}$ could then be calculated.

Python 2.7 was used as our main programming language, and the entire system was executed on a personal computer equipped with Intel Core i7-6800 3.4GHz CPU and 64GB memory. The processing time was counted from the input of an entire set of raw CT images till the MLS estimate was shown on the computer screen.

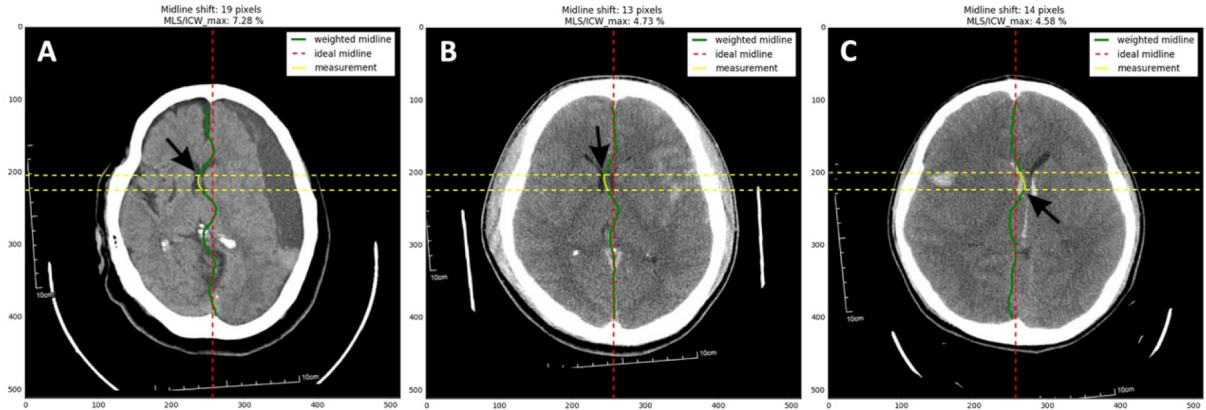

**Figure 3. MLS measurement by computer**
Three examples of computer measurements were shown here. WML (green solid line) was plotted by calculating weighted averages in each row in the image. The black arrows pointed out the point with maximal distance from the desired WML segment to the IML. MLS was calculated as the maximal distance between the desired WML segment near the foramen of Monro (yellow solid line) and the IML (red dotted line).
WML: weighted midline; IML: ideal midline; MLS: midline shift.

## RESULTS

*Within doctors:*

For the human-measured results, there was no statistical difference along with high degree of correlation between the results obtained from the 3 doctors. A score of 1 was designated if the answer to "first glance evaluation" was an "yes", and a score of 0 was given if the answer was "no"; summation of all three doctors' scores resulted in a total score of 0~3, a score of 2 or 3 was considered an agreement on significant MLS. This MLS agreement served as the gold standard (i.e. ground truth) of classification in our study for evaluation of accuracy. By plotting the receiver operating characteristic (ROC) curves, $MLS/ICW_{MAX}$ ratio (area-under-curve: 0.9935, Youden index: 0.9333) was found to be approximately equal to direct MLS measurement (area-under-curve: 0.9962, Youden index: 0.9615) in terms of MLS agreement. A threshold of 0.5 cm for determination of MLS was found to be not as good as a value around 0.38 cm in this particular dataset, as we evaluated the nearest point to the left upper corner on the ROC curve

**Table 2. Confusion matrices of MLS classification by doctors**

*Dr. 1 (threshold: 0.50 cm)*

| Pred / Act | MLS | No MLS |
|---|---|---|
| MLS | 12 | 3 |
| No MLS | 0 | 26 |

*Accuracy: 92.68%*

*Dr. 2 (threshold: 0.50 cm)*

| Pred / Act | MLS | No MLS |
|---|---|---|
| MLS | 10 | 5 |
| No MLS | 0 | 26 |

*Accuracy: 87.80%*

*Dr. 3 (threshold: 0.50 cm)*

| Pred / Act | MLS | No MLS |
|---|---|---|
| MLS | 11 | 4 |
| No MLS | 0 | 26 |

*Accuracy: 90.24%*

*Dr. 1 (threshold: 0.38 cm)*

| Pred / Act | MLS | No MLS |
|---|---|---|
| MLS | 15 | 0 |
| No MLS | 1 | 25 |

*Accuracy: 97.56%*

*Dr. 2 (threshold: 0.38 cm)*

| Pred / Act | MLS | No MLS |
|---|---|---|
| MLS | 13 | 2 |
| No MLS | 2 | 24 |

*Accuracy: 90.24%*

*Dr. 3 (threshold: 0.38 cm)*

| Pred / Act | MLS | No MLS |
|---|---|---|
| MLS | 14 | 1 |
| No MLS | 1 | 25 |

*Accuracy: 95.12%*

The classification results of doctors evaluated by confusion matrices. The upper three matrices came from data analyzed with MLS threshold set to 0.5 cm. The binary decision of MLS or no MLS was performed by each individual doctor's measurement results, and this decision was compared with the ground truth answer obtained from all three doctors' MLS agreement. The lower three matrices were generated with data analyzed in the same manner, only to change the threshold to 0.38 cm. Note that the overall accuracies became higher in all 3 doctors when the threshold was set to 0.38 cm.
MLS: midline shift; Pred: predicted classifications; Act: actual classifications.

**Table 3. Confusion matrices of MLS classification by computer**

*Manual calibration:*

*MLS (threshold: 0.50 cm)*

| Pred / Act | MLS | No MLS |
|---|---|---|
| MLS | 12 (80.0%) | 3 (20.0%) |
| No MLS | 2 (7.7%) | 24 (92.3%) |

*Accuracy: 87.80%*

*MLS/$ICW_{MAX}$ (threshold: 3.80%)*

| Pred / Act | MLS | No MLS |
|---|---|---|
| MLS | 12 (80.0%) | 3 (20.0%) |
| No MLS | 2 (7.7%) | 24 (92.3%) |

*Accuracy: 87.80%*

*MLS (threshold: 0.38 cm)*

| Pred / Act | MLS | No MLS |
|---|---|---|
| MLS | 14 (93.3%) | 1 (6.7%) |
| No MLS | 2 (7.7%) | 24 (92.3%) |

*Accuracy: 92.68%*

*MLS/$ICW_{MAX}$ (threshold: 2.90%)*

| Pred / Act | MLS | No MLS |
|---|---|---|
| MLS | 14 (93.3%) | 1 (6.7%) |
| No MLS | 2 (7.7%) | 24 (92.3%) |

*Accuracy: 92.68%*

*Automated calibration:*

*MLS (threshold: 0.50 cm)*

| Pred / Act | MLS | No MLS |
|---|---|---|
| MLS | 9 (60.0%) | 6 (40.0%) |
| No MLS | 1 (3.84%) | 25 (96.15%) |

*Accuracy: 82.93%*

*MLS/$ICW_{MAX}$ (threshold: 3.80%)*

| Pred / Act | MLS | No MLS |
|---|---|---|
| MLS | 9 (60.0%) | 6 (40.0%) |
| No MLS | 1 (3.84%) | 25 (96.15%) |

*Accuracy: 82.93%*

*MLS (threshold: 0.38 cm)*

| Pred / Act | MLS | No MLS |
|---|---|---|
| MLS | 12 (80.0%) | 3 (20.0%) |
| No MLS | 1 (3.84%) | 25 (96.15%) |

*Accuracy: 90.24%*

*MLS/$ICW_{MAX}$ (threshold: 2.90%)*

| Pred / Act | MLS | No MLS |
|---|---|---|
| MLS | 12 (80.0%) | 3 (20.0%) |
| No MLS | 1 (3.84%) | 25 (96.15%) |

*Accuracy: 90.24%*

The classification results of computer evaluated by confusion matrices. The upper four matrices came from data analyzed with MLS threshold set to 0.50 cm (or MLS/$ICW_{MAX}$ threshold set to 3.80%). The binary decision of MLS or no MLS was performed by computer's measurement results, and this decision was compared with the ground truth answer obtained from doctors' MLS agreement. The lower four matrices were generated from data analyzed in the same manner, only to change the MLS threshold to 0.38 cm (or MLS/$ICW_{MAX}$ threshold set to 2.90%). Matrices on the left were obtained from manually calibrated images plus automated measurements, and matrices on the right were obtained from automated calibration of head rotation and measurements.

The overall accuracies were 87.80% for manual calibration and 82.93% for automated calibration when MLS threshold of 0.50 cm (or MLS/$ICW_{MAX}$ threshold of 3.80%), and improved to 92.68% for manual calibration and 90.24% for automated calibration when MLS threshold of 0.38 cm (or MLS/$ICW_{MAX}$ threshold of 2.90%) was used instead. Note that the confusion matrices were identical between MLS threshold of 0.5 cm and MLS/$ICW_{MAX}$ threshold of 3.80%, as well as those between 0.38 cm MLS and 2.90% MLS/$ICW_{MAX}$.

MLS: midline shift; $ICW_{MAX}$: maximal intracranial width; Pred: predicted classifications; Act: actual classifications.

and the Youden index. A direct MLS measurement of 0.5 cm was approximately equivalent to a MLS/$ICW_{MAX}$ ratio of 3.80%, while a MLS of 0.38 cm was roughly equal to MLS/$ICW_{MAX}$ ratio of 2.90%. The performances of each doctor were shown in confusion matrices (Table 2), with overall accuracies ranged from 87.80% to 92.68% when a MLS threshold of 0.5 cm was used, and improved to 90.24% to 97.56% when a MLS threshold of 0.38 cm was used instead.

*Doctors v.s. computer:*

Although our automated system included a slice-selection algorithm to pick appropriate CT slice for MLS measurement, we designated pre-selected slices (and therefore did not deploy the automated slice-selection method in this study) in order to be certain that the human experts and the computer were seeing the same images. By using the standard MLS threshold of 0.5 cm, the performance was barely satisfactory with the overall accuracies of 87.80% by using manual calibrations of head rotation plus automated measurements, and 82.93% by a fully automated system. The performance was much improved by changing the MLS threshold to 0.38 cm, with the overall accuracies of 92.68% and 90.24% from manual calibrations and automated calibrations, respectively. These results were evaluated in confusion matrices as well (Table 3). Examples of automated MLS measurement are shown in Figure 3.

The ROC curves derived from individual doctor's results, average MLS measurement from all three doctors and the MLS measurement from computer all showed fairly large area-under-curve (Figure 4), and similar curves were also obtained from MLS/ICW$_{MAX}$ results. We also found a high degree of agreement among the computer measurements and the average manual measurements shown by the Bland-Altman difference plot, with the mean differences close to 0 and no obvious proportional bias observed (Figure 5).

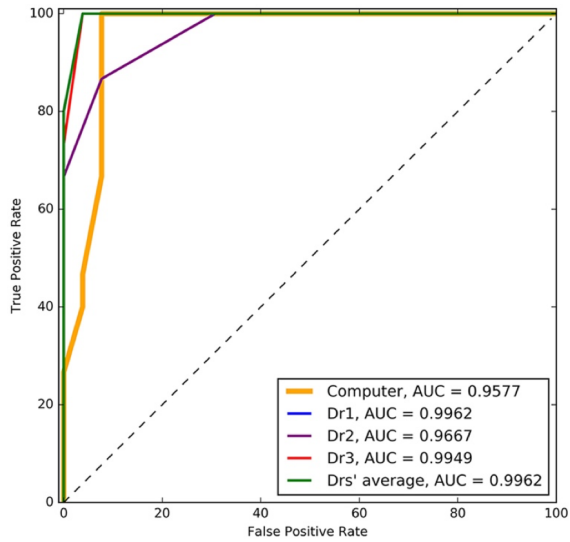

**Figure 4. ROC curves for computer and manual MLS measurements**
The ROC curves derived from individual doctor's results, average MLS measurement from all three doctors and the MLS measurement from computer, with all curves showing large AUC (0.9577 for computer, 0.9667~0.9962 for individual doctor and 0.9962 for average of all doctors). Results from MLS/ICW$_{MAX}$ showed similar results and we omitted them for redundancy.
ROC: receiver operating characteristic; MLS: midline shift; AUC: area-under-curve.

## DISCUSSION

*Fully automated system with universal applications:*

We have arrived in an era with emerging artificial intelligence and incorporation of smart hospital system, computer-aided diagnosis is one of the integral part that needs to be explored. Through the advancement of computational technologies, we believe that a fully automated system can have the potential of providing significant help to establish a fast and reliable diagnosis and also to decrease the burden of health care providers. To our understanding, the previously proposed systems are not really ready for clinical applications, as some of them did not state the inclusion criteria for CT image collection [5, 17] and therefore people may question whether their methods can be applied to other non-selected cases as well; others collected CT images in a standardized protocol [6, 7] but this is often not the clinical scenario. Human experts are trained to assess CT images obtained under different settings, therefore it is reasonable to expect that a capable automatic system can do the same. Our system is tested on a dataset consists of consecutively admitted patients under neurosurgical service, excluding only patients without CT images available in the PACS archive and two more other patients whose CT scans were difficult to read even for human experts, and the CT images were done under various settings as previously described (therefore resembles the real clinical situations).

*Processing speed:*

Most of the previous semi-automated computer-aided diagnosis involving measurement of MLS on brain CT images focused on their accuracies without much mention of the processing speed [5-7, 17], however, the high-level computer vision techniques such as template matching for the ventricles can be very time-consuming. As stated earlier, our method of deciding IML is simple and robust. The overall processing speed of MLS measurement for our system is quite fast, which only takes about 10 seconds for a set of images (around 30 to 40 axial CT slices) to be processed.

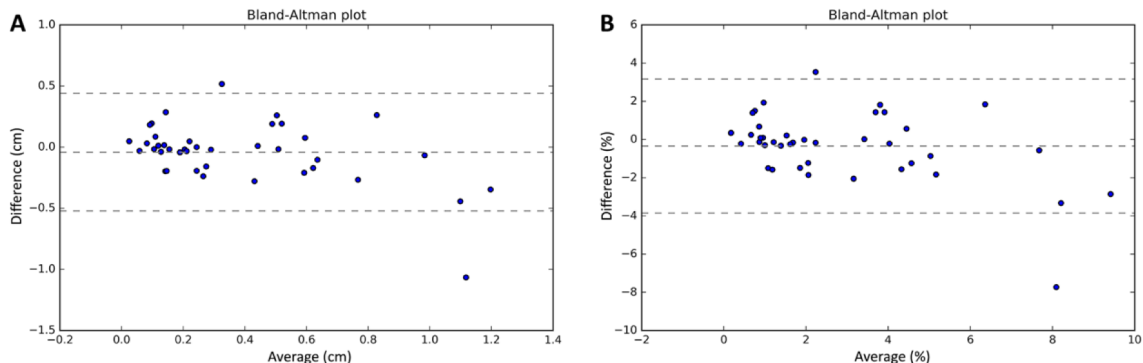

**Figure 5. Bland-Altman plots for computer *vs* manual MLS measurements**
Bland-Altman plots generated from computer measurements and the average of manual measurements. A: plot with results of MLS; B: plot with results of MLS/ICW$_{MAX}$. The dotted lines represented the mean difference (line in the middle) and the boundaries of 1.96 times the standard deviation of the differences, which marked the 95% confidence interval. The mean difference was close to 0 in both plots, and no obvious proportional bias was observed.

*Accurate estimation:*

One of the most important metric in daily medical practice is the false negative rate, since a wrong diagnosis of no MLS might potentially delay the treatment of an emergent patient. Our system made only one out of fifteen for this false negative mistake if the CT scan was manually calibrated, and the false negatives rose to three cases if the CT scan was automatically calibrated. Either condition was comparable to the performance of human experts, as shown in Table 2 and Table 3. In addition to this low false negative rate, an overall accuracy of 90.24% and 92.68% was demonstrated if the calibration process was done automatically and manually, respectively. Both results are also comparable to human experts (who performed 97.56%, 90.24% and 95.12%, respectively). Comparing this result to previous publications, an overall accuracy of 76.54% (62 out of 81 cases) was reported by Liao et al. while Chen et al. [5, 14] and Liu et al. [17] did not report their overall accuracy on the results of MLS estimation.

It is difficult to determine the ground truth of MLS in each CT image set because results from the measurements of experts still vary. Therefore, we adopted agreement from the majority vote of "first glance evaluation" as the answer for ground truth. This first glance evaluation was considered to be of significance because trained neurosurgeons perceived the entire set of CT images as a whole patient, and evaluated the mass effect not just by MLS measurement but rather in a global perspective. The length of MLS is just a simplified and quantified parameter of the mass effect. We asked the human experts to answer whether there was a significant MLS or not, but the response will definitely include some of their judgment from the observations on the mass effect as a whole, and will better reflect the conditions of brain deformity. To our surprise, the performance of MLS threshold at 0.38 cm surpassed that at 0.5 cm in our dataset (even though all of our human experts are trained to think that a MLS greater than 0.5 cm is significant). This cannot be taken as a standard to apply universally because the threshold value should be data-dependent which may not work so well in other datasets. Larger dataset will be tested in order to find the optimal threshold that fits to all circumstances.

Previous publications emphasized on how to plot the whole deformed midline [5-7, 17]. While the standard measurement of MLS only uses the septum pellucidum at the level of foramen of Monro, we find it redundant to depict the rest of the deformed midline. As we put more stress over the frontal horns of the lateral ventricles, we found it extremely simple to track just the dark portions inside the intracranial region. Due to this simplicity, the algorithm can be easily generalized for more complicated cases. For instance, the original algorithm was modified such that the MLS in cases with intraventricular hemorrhage (IVH) or missing ventricles can still be detected (Figure 3C).

*$MLS/ICW_{MAX}$ ratio:*

Our last remark is to emphasize the ratio of $MLS/ICW_{MAX}$ proposed in this study. We initially proposed the use of this ratio based on the question that whether an arbitrary absolute threshold value of MLS should be applied to all patients, especially when considering different age groups and different ethnicities. By using this ratio, we found that it was also very useful to overcome the generic difficulty of turning the measured pixels into centimeters when only non-DICOM images (such as JPEG files in our case) were available, of which the information of pixel spacing was missing. We demonstrated that this $MLS/ICW_{MAX}$ ratio was a nice alternative for MLS estimation, especially when we are aiming to develop a fully automated computer-aided system.

## CONCLUSION

We present a simple and effective way to detect the MLS, which is essentially tracking the dark portions of brain tissues in axial brain CT slices. These dark portions of brain tissues in the appropriately selected slices are essentially ventricles, which are the reference points for human experts to determine MLS in individual cases. The $MLS/ICW_{MAX}$ ratio was found to be highly correlated to direct MLS measurement and can be used as an alternative choice of evaluating the mass effect in brain CT images. A ratio rather than an absolute value also expands its potential use, especially when non-DICOM images (e.g. JPEG files) without known information on pixel spacing are used. In this study, we have demonstrated the feasibility of a fast and fully automated approach for MLS measurement on brain CT images with satisfactory accuracy.

## ACKNOWLEDEMENT

We thank Dr. Chien-Hsun Li at National Taiwan University Hospital Hsin-Chu Branch and Dr. Kuo-Wei Chen at National Taiwan University Hospital for the measurement of brain CT parameters. There is no conflict of interest to declare.